\documentclass[aip,apl,floatfix,reprint]{revtex4-1}
\usepackage[colorlinks=true,citecolor=blue,menucolor=blue,linkcolor=blue,urlcolor=blue]{hyperref}

\usepackage{amsfonts}
\usepackage{amssymb}
\usepackage{amsmath}
\usepackage{longtable}
\usepackage{graphicx}
\usepackage[tabke]{xcolor}
\usepackage{multirow}

\begin{document}

\title{Optical vs electronic gap of hafnia by ab initio Bethe-Salpeter equation}

\author{Beno\^{i}t Skl\'{e}nard}
\email{benoit.sklenard@cea.fr}
\affiliation{Univ. Grenoble Alpes, F-38000 Grenoble, France}
\affiliation{CEA, LETI, MINATEC Campus, F-38054 Grenoble, France}

\author{Alberto Dragoni}
\affiliation{Univ. Grenoble Alpes, F-38000 Grenoble, France}
\affiliation{CEA, LETI, MINATEC Campus, F-38054 Grenoble, France}
\affiliation{CNRS, Institut N\'{e}el, F-38042 Grenoble, France}

\author{Fran\c{c}ois Triozon}
\affiliation{Univ. Grenoble Alpes, F-38000 Grenoble, France}
\affiliation{CEA, LETI, MINATEC Campus, F-38054 Grenoble, France}

\author{Valerio Olevano}
\affiliation{Univ. Grenoble Alpes, F-38000 Grenoble, France}
\affiliation{CNRS, Institut N\'{e}el, F-38042 Grenoble, France}

\date{\today}

\begin{abstract}
We present first-principles many-body perturbation theory calculations of the quasiparticle electronic structure and of the optical response of HfO$_2$ polymorphs.
We use the $GW$ approximation including core electrons by the projector augmented wave (PAW) method and performing a quasiparticle self-consistency also on wavefunctions (QS$GW$).
In addition, we solve the Bethe-Salpeter equation on top of $GW$ to calculate optical properties including excitonic effects.
For monoclinic HfO$_2$ we find a fundamental band gap of $E_g = 6.33$~eV (with the direct band gap at $E_g^d = 6.41$~eV), and an exciton binding energy of 0.57~eV, which situates the optical gap at $E^o_g = 5.85$~eV.
The latter is in the range of spectroscopic ellipsometry (SE) experimental estimates (5.5-6~eV), whereas our electronic band gap is well beyond experimental photoemission (PE) estimates ($< 6$~eV) and previous $GW$ works.
Our calculated density of states and optical absorption spectra compare well to raw PE and SE spectra. This suggests that our predictions of both optical and electronic gaps are close to, or at least lower bounds of, the real values.
\end{abstract}

\maketitle

\paragraph*{Introduction}
Hafnia (HfO$_2$) is a transition metal oxide having attracted much attention due to its numerous technological applications, mainly related to its optical and electrical insulating properties. 
It is used for optical coatings\cite{Gilo_1999} in the near-ultraviolet (UV) to infrared (IR) wavelengths range, or as a high permittivity dielectric\cite{Robertson_2006} in submicrometer silicon-based technologies.
More recently, it is gaining interest as an insulating layer in resistive random access memories (ReRAM)\cite{Wong_2012,Bersuker_2011} which are a promising candidate for the next-generation nonvolatile memories. 

At ambient pressure, bulk HfO$_2$ exists in three thermodynamically stable crystalline polymorphs. At low temperature, the most stable phase exhibits a monoclinic $P2_1/c$ symmetry (m-HfO$_2$), and transforms into a $P4_2/nmc$ tetragonal phase (t-HfO$_2$) around 2000~K.\cite{Wolten_JACS_1963} At higher temperature, the tetragonal structure undergoes another phase transition to a $Fm3m$ cubic fluorite symmetry (c-HfO$_2$). 
In contrast to bulk samples, as-deposited HfO$_2$ thin films are typically amorphous but crystallize after anneal.\cite{Ho_JAP_2003} 
After crystallization, the lowest-energy monoclinic phase is prevalent, but the presence of tetragonal and metastable orthorhombic phases have also been observed.\cite{Ho_JAP_2003,Nguyen_APL_2005,Park_JAP_2008,Hill_JAP_2008} 
The monoclinic is therefore the reference phase and we will mainly focus on it, except when the other phases are explicitly mentioned.

Several experimental techniques have been employed to characterize HfO$_2$ thin films.
On the one hand, both X-ray (XPS) and ultraviolet (UPS) photoemission spectroscopy (PES), and inverse photoemission (IPS) have been used to study the electronic structure. \cite{Sayan_JAP_2004,Bersch_PRB_2008}
On the other hand, X-ray or optical absorption, \cite{Balog_TSF_1977,Toyoda_JESRP_2004} spectroscopic ellipsometry (SE),\cite{Zhu_IEEEEDL_2002,Edwards_AIPCP_2003,Modreanu_PSPIE_2003,Nguyen_APL_2005,Park_JAP_2008,Hill_JAP_2008} and electron energy-loss spectroscopy (EELS) \cite{Yu_APL_2002,Puthenkovilakam_JAP_2004,Cheynet_JAP_2007,Ikarashi_JAP_2003,Guedj_APL_2014,Hung_PRB_2016} have been used to study the optical and dielectric properties.
By using linearization and extrapolation techniques over measured spectra, these experiments extracted gap values ranging from 5.1\cite{Toyoda_JESRP_2004} to 5.95 eV\cite{Modreanu_PSPIE_2003}.
Surprisingly, the ranges for optical (5.1--5.95~eV) and electronic gaps (5.7--5.86~eV)\cite{Sayan_JAP_2004,Bersch_PRB_2008} overlap, making unclear the distinction between them.

The electronic structure of HfO$_2$ polymorphs has also been studied theoretically.
First works on m-HfO$_2$~\cite{Boeri_JPCM_1998,Demkov_PSSB_2001,Peacock_JAP_2002,Zhao_PRB_2002,Fiorentini_PRL_2002}, by using density functional theory (DFT) in the local density (LDA) or generalized gradient approximation (GGA)\cite{PBE96}, found band gaps (3.8--4.0~eV) underestimated by 30\% with respect to experimental data.
Most recent works, by using advanced semi-empirical functionals like TBmBJ\cite{Ondracka_JPD_2016} or many-body perturbation theory (MBPT) in the $GW$ approximation~\cite{Gruning_PRB_2010,Jiang_PRB_2010,VanDerGeest_PRB_2012}, found band gaps in the range 5.7--5.9~eV and reconciled a good agreement with experimental data. These studies do not account for excitonic effects but agree well with optical gaps derived from SE and EELS.  

In this work we revisit the situation.
We calculate the electronic structure in the framework of MBPT within the $GW$ approximation,\cite{Hedin65,StrinatiHanke80,StrinatiHanke82} also including core electrons by the projector-augmented wave (PAW) method \cite{Blochl_PRB_1994} and applying self-consistency on wavefunctions  (QS$GW$).\cite{Faleev_PRL_2004} On top of QS$GW$, we perform Bethe-Salpeter equation (BSE) calculations \cite{SalpeterBethe51,HankeSham74,HankeSham79} of the optical gap and spectra including electron-hole interaction (excitonic effects). Furthermore, we perform a careful convergence study of our results (See supplementary material). Finally, instead of comparing our gaps with experimental values, we compare our DOS and optical absorption directly with the raw measured spectra.
Our results indicate 5.85~eV (the energy of the first exciton) as a lower bound for the the m-HfO$_2$ optical gap.
This is still in the range of the experimentally derived optical gaps.
On the other hand, the comparison of our DOS with PES spectra clearly indicate 6.33~eV as a lower bound for the  m-HfO$_2$ electronic band gap.
Our BSE calculation indicates the first peak of optical absorption as due to an exciton whose binding energy is 0.57~eV.

\begin{figure}
\centering
\includegraphics[width=\linewidth]{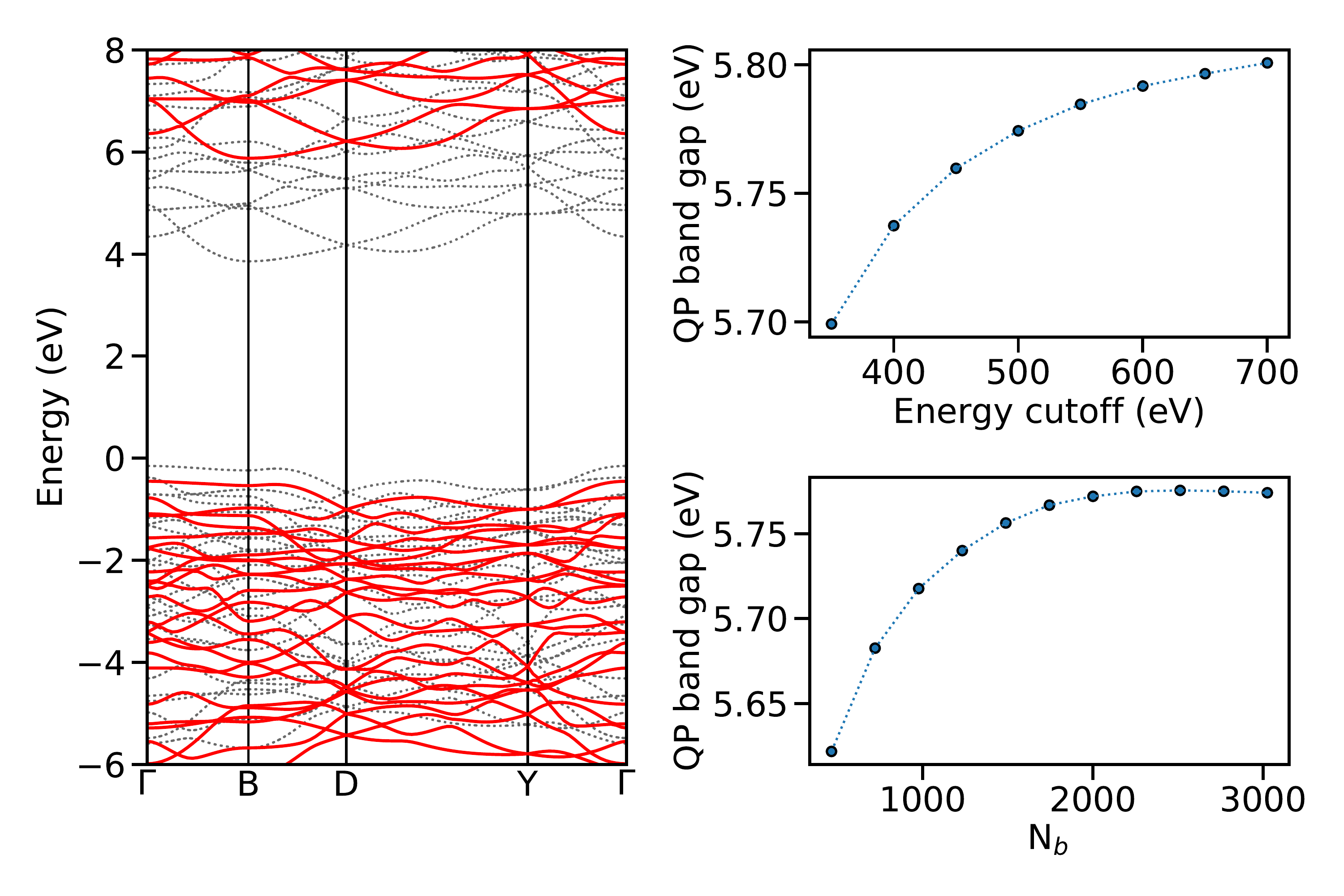}
\caption{
Left: Calculated band structure of monoclinic HfO$_2$ within the DFT-PBE (dotted line) and QS$GW$ (plain red line) approximations.
The QS$GW$ band structure has been interpolated using MLWF.
PBE and $GW$ Fermi energies are aligned at zero.
Top right: convergence of the QP gap at $\Gamma$ at the $G_0W_0$ level as a function of plane wave energy cutoff, with N$_b$ fixed to the maximum number of plane waves. Bottom right: convergence of the QP gap at $\Gamma$ at the $G_0W_0$ level as a function of the number of empty bands included (N$_b$) with a 500~eV plane wave energy cutoff.
}
\label{band-plot}
\end{figure}

\paragraph*{Computational details}
\textit{Ab initio} calculations based on density functional theory\cite{HohenbergKohn64,KohnSham65} in the LDA or PBE\cite{PBE96} approximations are carried out using the \textsc{vasp} code.\cite{Kresse_PRB_1996, Kresse_PRB_1999} 
The core-valence interaction are described with PAW datasets including the semicore $5s$ and $5p$ states for Hf.
Electron wave functions are expanded in a plane waves basis set with kinetic energy cutoff of 500~eV, and the Brillouin zone is sampled using $4 \times 4 \times 4$, $6 \times 6 \times 4$ and $6 \times 6 \times 6$ $\Gamma$-centered Monkhorst-Pack meshes for monoclinic,  tetragonal and cubic HfO$_2$ phases, respectively.
Many-body effects are accounted for by computing the quasiparticle (QP) energies at the $G_0W_0$, $GW_0$ (self-consistency only on the eigenvalues) and QS$GW$ level on top of DFT but fixing $W$ at $W_0$. Indeed, fixing $W$ has been shown to improve the agreement with experimental band gaps than using self-consistent $W$.\cite{Shishkin_PRB_2007}
In contrast to $G_0W_0$ and $GW_0$, QS$GW$ allows to reduce the influence of the DFT starting point (LDA vs. PBE) on the electronic structure.
The cutoff for response function is taken to be 333~eV and about 500 empty bands per formula unit and 100 frequency grid points are needed to obtain converged band gaps within 0.1~eV (see supplementary material for details). 
QP band structures are interpolated using maximally localized Wannier functions (MLWF) with the \textsc{Wannier90} code.\cite{Mostofi_CPC_2014}
To determine the optical properties, the Bethe-Salpeter equation (BSE) is solved on top of $GW$ using the Tamm-Dancoff approximation.\cite{Sander_PRB_2015} 

\paragraph*{Band gap and electronic structure}
In Fig.~\ref{band-plot} we report the electronic band structure of m-HfO$_2$, as calculated within DFT in the PBE approximation and within QS$GW$.
In the monoclinic crystal structure HfO$_2$ presents a direct band gap at $B$, whereas the fundamental minimum band gap is indirect at $\Gamma \to B$.
Between DFT and the $GW$ approximation there is some rearrangement of the bands, but the most important effect is a shift of both valence and conduction bands which increases band gaps.
This is also what we found for the cubic and tetragonal phases.

\begin{table}[b]
\caption{
Fundamental minimum band gap for the cubic ($X \rightarrow X$), tetragonal ($Z \rightarrow \Gamma$), and monoclinic ($\Gamma \rightarrow B$)  HfO$_2$ phases, as calculated in various approximations.
Ref.~\onlinecite{Gruning_PRB_2010} used a norm-conserving pseudopotentials plane waves approach, whereas Ref.~\onlinecite{Jiang_PRB_2010} and Ref.~\onlinecite{Ondracka_JPD_2016} used FP-LAPW.
The determination of the band gap from PES/IPS experiments relies on post-processing data analysis to remove tails due to impurities and fits: for the XPS+IPS experiment of Ref.~\onlinecite{Sayan_JAP_2004} we quote two estimates removing or not the effect of tails.
}
\begin{ruledtabular}
\begin{tabular}{lccc}
 method   & \phantom{te}cubic\phantom{te} & tetragonal & monoclinic \\
 \hline
 \multicolumn{4}{c}{This work} \\
 DFT-LDA      & 3.68 & 4.41 & 3.93 \\
 LDA+$G_0W_0$ & 5.63 & 6.40 & 5.95 \\
 LDA+$GW_0$   & 6.01 & 6.79 & 6.36 \\
 LDA+QS$GW$   & 6.14 & 6.89 & 6.37 \\
 DFT-PBE      & 3.76 & 4.64 & 4.01 \\
 PBE+$G_0W_0$ & 5.41 & 6.34 & 5.77 \\
 PBE+$GW_0$   & 5.78 & 6.72 & 6.18 \\
 PBE+QS$GW$   & 6.03 & 6.93 & 6.33 \\
 \hline
 \multicolumn{4}{c}{Other theoretical works} \\
 DFT-LDA\cite{Gruning_PRB_2010} & 3.5 & 4.1 & 3.8 \\
 LDA+$G_0W_0$\cite{Gruning_PRB_2010,Gruning_private_2018} & 5.2 & 5.8 & 5.7 \\
 LDA+$GW_0$\cite{Gruning_PRB_2010} & 5.5 & 6.0 & 5.9 \\
 DFT-LDA\cite{Jiang_PRB_2010} & 3.55 & 4.36 & 3.95 \\
 LDA+$G_0W_0$\cite{Jiang_PRB_2010} & 4.91 & 5.78 & 5.45 \\
 LDA+$GW_0$\cite{Jiang_PRB_2010} & 5.20 & 6.11 & 5.78 \\
 DFT-PBE\cite{Ondracka_JPD_2016} & 3.77 & 4.79 & 4.08 \\
 TB-mBJ orig\cite{Ondracka_JPD_2016} & 5.88 & 6.54 & 5.76 \\
 TB-mBJ imp\cite{Ondracka_JPD_2016} & 6.17 & 6.81 & 6.01 \\
 TB-mBJ semi\cite{Ondracka_JPD_2016} & 6.74 & 7.35 & 6.54 \\
 \hline
 \multicolumn{4}{c}{Experimental works} \\
 \multicolumn{3}{l}{UPS+IPS (straight line extrapolation)\cite{Bersch_PRB_2008}} & 5.7 \\
 \multicolumn{3}{l}{XPS+IPS (straight line extrapolation)\cite{Sayan_JAP_2004}} & 5.86 \\
 \multicolumn{3}{l}{XPS+IPS (comparison with shifted DFT)\cite{Sayan_JAP_2004}} & 6.7 \\
\end{tabular}
\end{ruledtabular}
\label{band gaps}
\end{table}

In Table~\ref{band gaps} we report the fundamental minimum gaps for all phases and approximations considered in this work, and we compare them to previous theoretical works and to experiments.
Regarding DFT results, our PAW LDA and PBE gaps are more in agreement with the FP-LAPW (full potential linearized augmented plane wave) LDA and PBE gaps of, respectively, Ref.~\onlinecite{Jiang_PRB_2010} and \onlinecite{Ondracka_JPD_2016} (both are all-electron calculations), than with the norm-conserving pseudopotentials plane waves (NCPP PW) LDA gap of Ref.~\onlinecite{Gruning_PRB_2010}.
\begin{figure}[t]
\includegraphics[width=\linewidth]{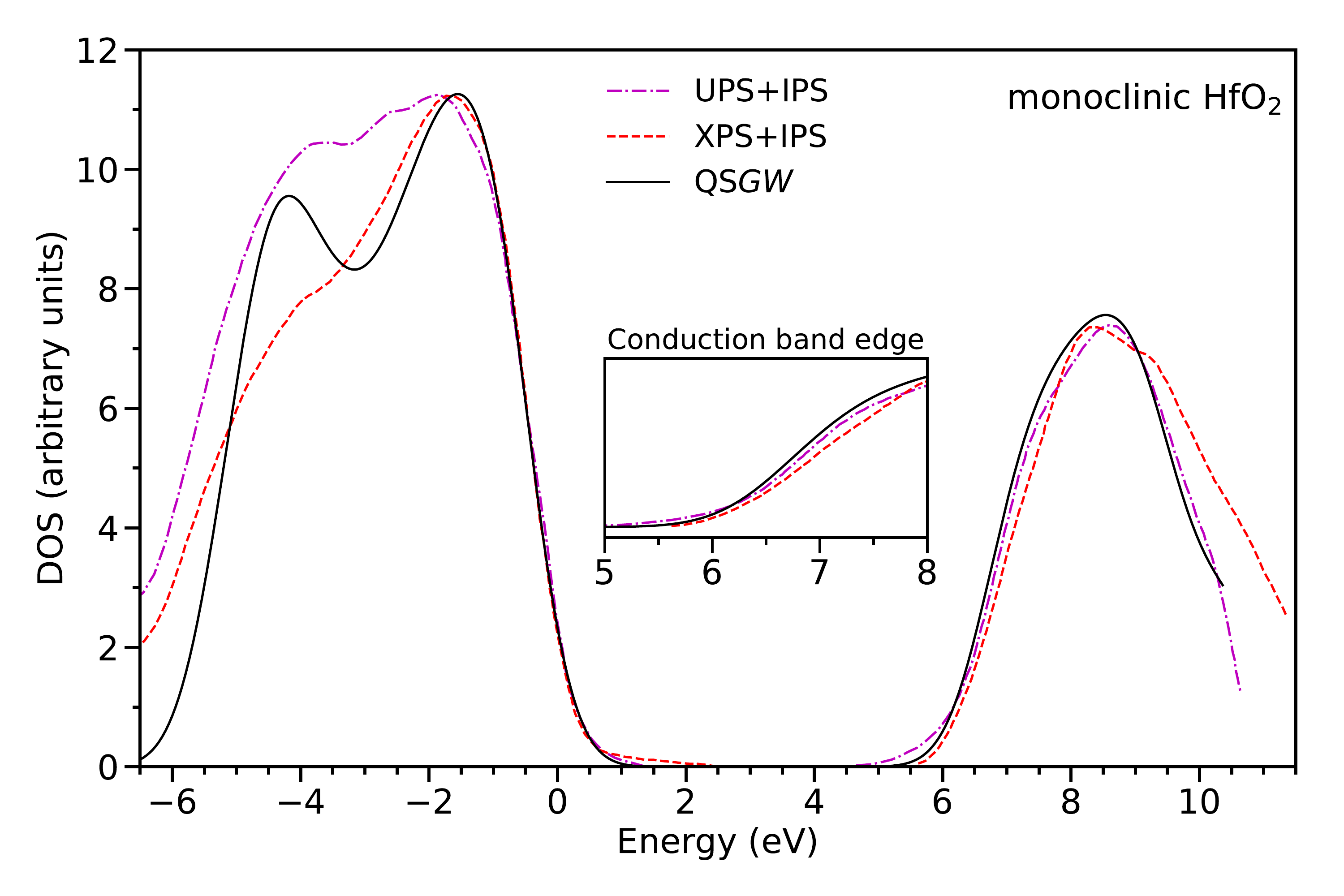}
\caption{
Density of states (DOS) of monoclinic HfO$_2$ in the QS$GW$ approximation compared to UPS+IPS\cite{Bersch_PRB_2008} and XPS+IPS\cite{Sayan_JAP_2004} spectra.
The QS$GW$ DOS has been interpolated using MLWF on a $40 \times 40 \times 40$ k-point grid and convoluted with a gaussian broadening of 0.7~eV.
Experimental spectra have been aligned at the valence band-edge (the zero of the energy for the theoretical DOS), and the conduction IPS and valence PS independent measurements have been rescaled separately to match the height of the theoretical DOS.
The error in this procedure can be estimated by the maximum deviation among valence band-edges in the steepest rising linear region: 50~meV, which is less than our theoretical error bar.
}
\label{dos}
\end{figure}
On the other hand, both our $G_0W_0$ and $GW_0$ band gaps are systematically larger than the ones of Refs.~\onlinecite{Gruning_PRB_2010,Jiang_PRB_2010}.
We remark however that our PAW $G_0W_0$ corrections, $\Delta E_g^{GW} = E_g^{GW} - E_g^\mathrm{DFT}$, are closer to the NCPP PW ones\cite{Gruning_PRB_2010} than to the FP-LAPW\cite{Jiang_PRB_2010} ones, which are 0.3 to 0.6~eV lower.
Differences between Ref.~\onlinecite{Gruning_PRB_2010} and our $G_0W_0$ gaps could then be explained by the different starting DFT gap.
Application of self-consistency only on eigenvalues, using the $GW_0$ approach, further increases the gap by $\sim$0.4~eV in our study and by 0.2$\sim$0.3~eV in Refs.~\onlinecite{Gruning_PRB_2010,Jiang_PRB_2010}.
Full self-consistency also on the wavefunctions,\cite{Rangel_PRB_2011} using the QS$GW$ approach,\cite{Faleev_PRL_2004} removes any influence of the LDA or PBE starting point in our study, reducing the gap difference to less than 0.1~eV, 
a residual due to the use of different relaxed LDA and PBE atomic structures.

In the following we consider only PBE relaxed atomic structure, the closest to the experiment.
Our $GW_0$ and QS$GW$ calculations systematically yield larger band gaps than previous theoretical studies.
For the monoclinic phase, our values are 0.4$\sim$0.5~eV larger than the highest $GW_0$ estimates of Refs.~\onlinecite{Jiang_PRB_2010,Gruning_PRB_2010} and the TB-mBJ-orig of Ref.~\onlinecite{Ondracka_JPD_2016}.
The latter are in very good agreement with the 5.7 and 5.86~eV band gap values determined from photoemission experiments,\cite{Bersch_PRB_2008,Sayan_JAP_2004} whereas our QS$GW$ gap of 6.33~eV appears as a large overestimation.
However, as discussed in Secs~III.C and F of Ref.~\onlinecite{Sayan_JAP_2004}, there is some uncertainty in this determination of band gaps by the conventional method of linear extrapolation of photoemission band edges to the background intensity, due to the presence of band-tail and defects in the vicinity of the valence band maximum and conduction band minimum. 
For this reason we prefer to directly compare the experimental PES+IPS spectra to our calculated DOS (Fig.~\ref{dos}).
This comparison was already suggested in the same Ref.~\onlinecite{Sayan_JAP_2004} to provide a safer estimate of the real band gap.
By using a DFT-LDA DOS and evaluating the scissor operator shift to make theoretical and experimental DOS coincide, they arrived to an estimate of 6.7~eV for the m-HfO$_2$ band gap\cite{Sayan_JAP_2004}.
Our QS$GW$ DOS favorably compares with photoemission spectra, especially on the shape, even though we have not taken into account extrinsic and finite state effects which are evident when comparing XPS with UPS shapes.
As it can be estimated by the deviation of theory and experiment in the conduction edge (Fig.~\ref{dos} inset), our QS$GW$ band gap of 6.33~eV is still an underestimation of about 0.2~eV of the real band gap.
Our more prudent conclusion is that the real band gap of monoclinic HfO$_2$ is $E_g > 6.33$~eV, and probably $E_g = 6.5$~eV.
This is also what Ondra\v cka et al.\cite{Ondracka_JPD_2016} found when modifying the TB-mBJ functional (``semi'' version in Table~\ref{band gaps}) to target the experimental DOS.
The QS$GW$ approximation has been reported to systematically overestimate band gaps in all studied materials.~\cite{Bruneval_2014}
In our case, for m-HfO$_2$, the close agreement between QS$GW$ and experimental spectra may be due to  fortuitous error cancellation  with other effects not taken into account, such as electron-phonon,\cite{e-ph} and both single-particle (e.g.\ spin-orbit) or many-body (e.g.\ Breit interaction) relativistic corrections.

\begin{table}[b]

\caption{
Direct electronic band gap (DFT or $GW$) and optical gap (BSE first exciton eigenvalue energy) for the cubic ($X \rightarrow X$), tetragonal ($\Gamma \rightarrow \Gamma$), and monoclinic ($B \rightarrow B$)  HfO$_2$ phases. We report also the exciton binding energy equal to the difference between the QS$GW$ direct band gap and the BSE optical gap, $E^{exc}_b = E^{d}_g - E^{o}_g$. For m-HfO$_2$, BSE gap agrees well with SE (spectroscopic ellipsometry) optical onset.
Energies are in eV. 
}
\begin{ruledtabular}
\begin{tabular}{lccc}
 method   & \phantom{te}cubic\phantom{te} & tetragonal & monoclinic \\
 \hline
 \multicolumn{4}{c}{This work}      \\
 DFT-LDA         & 3.68 & 4.58 & 4.03         \\
 LDA+$G_0W_0$    & 5.63 & 6.57 & 6.05         \\
 LDA+$GW_0$      & 6.01 & 6.96 & 6.46         \\
 LDA+QS$GW$      & 6.14 & 7.04 & 6.47         \\
 DFT-PBE         & 3.76 & 4.71 & 4.09 \\
 PBE+$G_0W_0$    & 5.41 & 6.43 & 5.86 \\
 PBE+$GW_0$      & 5.78 & 6.81 & 6.27 \\
 PBE+QS$GW$      & 6.03 & 7.01 & 6.41 \\
 PBE+QS$GW$+BSE  & 5.57 & 6.53 & 5.85 \\
 \hline
 $E^\mathrm{exc}_b$  & 0.46 & 0.48 & 0.57 \\
 \hline 
 \multicolumn{4}{c}{Experimental works (optical gap)} \\
 \multicolumn{3}{l}{SE\cite{Nguyen_APL_2005}} & 5.6--5.8 \\
 \multicolumn{3}{l}{SE\cite{Hill_JAP_2008}} & 5.5--6.0 \\
\end{tabular}
\end{ruledtabular}
\label{opticalgap}
\end{table}

\begin{figure}[t]
\includegraphics[width=\linewidth]{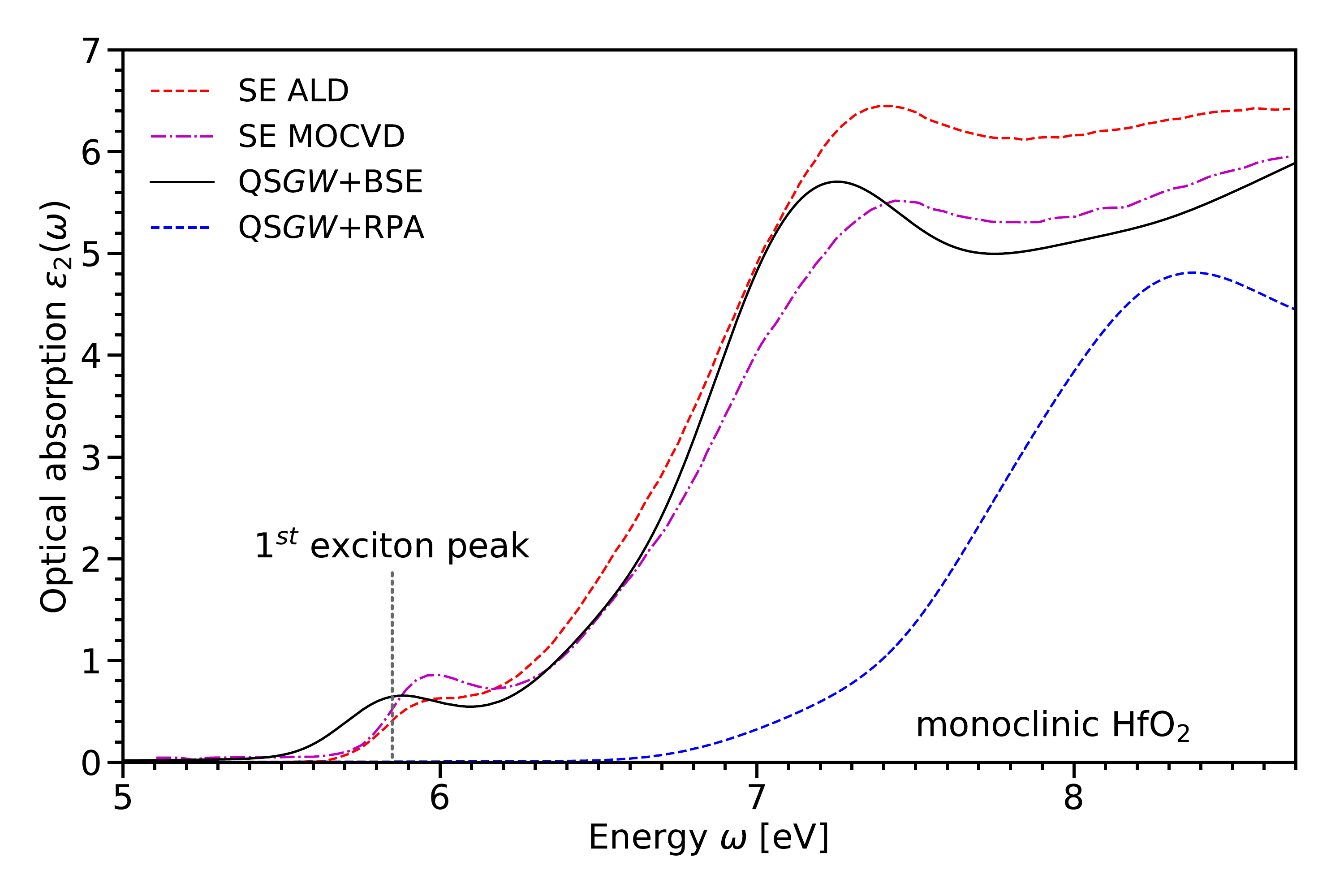}
\caption{
Imaginary part of the dielectric function $\varepsilon_2$ averaged over the three monoclinic HfO$_2$ lattice directions and convoluted with a gaussian broadening of 0.2 eV, compared to ellipsometry spectra (SE)\cite{Hill_JAP_2008}.
}
\label{bse} 
\end{figure}

\paragraph*{Optical gap and spectra}
In Table~\ref{opticalgap} we report all the DFT and $GW$ direct band gaps, and we add the optical gaps calculated by solving the BSE on top of QS$GW$.
We also report the first exciton binding energy, defined as the difference between the direct band gap energy and the energy of the first exciton, $E_b^\mathrm{exc} = E_g^d - E_g^o$, found to be 0.57 eV for m-HfO$_2$.
Direct band gap and optical gap are significantly different in m-HfO$_2$.
The simulated optical gap can now be compared with the measured one, e.g., in optical or X-ray absorption, spectroscopic ellipsometry (SE), or energy-loss (EELS).
Refs.~\onlinecite{Nguyen_APL_2005,Hill_JAP_2008} reported values derived from SE spectra of 5.6-5.8~eV and 5.5-6.0~eV, respectively.
The uncertainty is due, as for the band edges in the density of states, to the method (e.g.\ Tauc-Lorentz) used to linearly extrapolate to the background.
However, with respect to photoemission, optical experiments are less affected by defect, surface, interface or substrate effects and more sensitive to the bulk.
We remark that now our QS$GW$+BSE optical gap is in the range of the experimental reports.
Nevertheless, we again\cite{expabinitiogapcomp} prefer to compare the raw SE spectra to our calculated optical absorption, $\varepsilon_2(\omega)$ (see Fig.~\ref{bse}).
The QS$GW$+BSE dielectric function significantly improves the lower level of approximation QS$GW$+RPA, and achieves a very good agreement with SE spectra.
BSE introduces electron-hole interaction effects and gives rise to the exciton which is to be identified with the first peak of the BSE spectrum, absent in the RPA.
Nevertheless we remark a 0.1$\sim$0.2~eV red shift of the exciton peak with respect to its position in SE spectra.
Part of this red shift could be corrected by $k$-point sampling extrapolation to zero\cite{FuchBechstedt_2008}.
Nevertheless, like the band gap, also our optical gap suffer an underestimation, so that they have both to be regarded as lower bounds of the real values.

\begin{figure}[t]
\includegraphics[width=\linewidth]{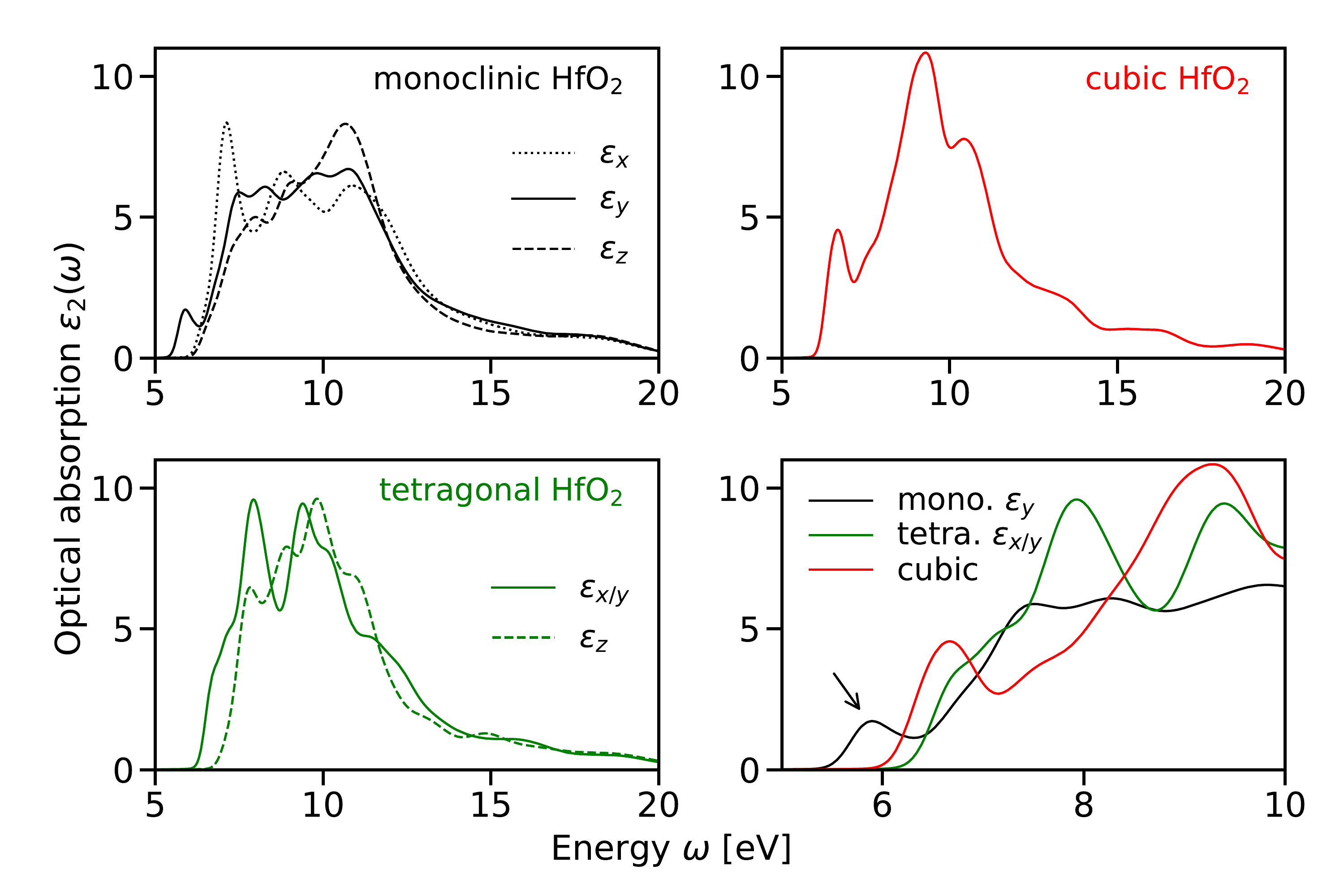}
\caption{
Imaginary part of the dielectric function $\varepsilon_2$ for the three HfO$_2$ phases and all, (100), (010), and (001) polarization directions.
The arrow indicates the only non-dark bound exciton in the (010) polarization of m-HfO$_2$.
}
\label{mtccomp} 
\end{figure}

The nature of this small peak at the optical onset observed in SE spectra\cite{Nguyen_APL_2005,Park_JAP_2008,Hill_JAP_2008} of crystalline samples has been attributed to different causes.
By combining X-ray absorption (XAS), X-ray diffraction (XRD) and SE techniques, Hill~\textit{et al.}\cite{Hill_JAP_2008} found that this feature could be intrinsic to the monoclinic phase.
According to our analysis, this is a real bound\cite{boundexcitons} exciton peak, as correctly interpreted in Refs.~\onlinecite{Edwards_AIPCP_2003,Ondracka_JPD_2016}, and not a defect state, as interpreted by Nguyen et al.\cite{Nguyen_APL_2005}.
The exciton is present only in the $y$ polarization (see Fig.~\ref{mtccomp}), confirming the unusual anisotropy in the dielectric properties of m-HfO$_2$\cite{Guedj_APL_2014,Hung_PRB_2016,anisotropy}.
We found an exciton also in c-HfO$_2$ and t-HfO$_2$, but their oscillator strength is zero or almost, so that they are dark excitons not detectable in SE spectra.
Hence SE spectra can be used to characterize the HfO$_2$ monoclinic phase with respect to all other phases by simply detecting the presence or absence of the 5.85~eV exciton peak.

\paragraph*{Conclusions} 
In this work we combine QS$GW$ calculations to compute the electronic structure of HfO$_2$ with BSE to compute optical spectra. We compare our calculated DOS and optical absorption with raw, as-acquired, experimental spectra measured for the monoclinic phase. Our calculated electronic band gap ($E_g = 6.33$~eV) is significantly larger than the values obtained in previous theoretical and experimental studies. However the direct comparison between QS$GW$ DOS and experimental spectra shows a good agreement and even indicates that our band gap value slightly underestimates by $\sim0.2$~eV the real value.
In contrast, we obtain an optical gap of 5.85~eV, in agreement with SE estimates. We find that the difference is due to the presence of a bound exciton with a large binding energy of 0.57~eV.

\paragraph*{Supplementary Material}
See supplementary material for information about the relaxed structures and convergence studies of $GW$ and BSE calculations.

\paragraph*{Acknowledgements}
Part of the calculations were run on TGCC/Curie using allocations from GENCI.

\begin{table*}[t!]
\renewcommand{\tabcolsep}{5mm}
\begin{ruledtabular}
\begin{tabular}{lcccccc}
 & \multicolumn{2}{c}{cubic} & \multicolumn{2}{c}{tetragonal} & \multicolumn{2}{c}{monoclinic}  \\ \cline{2-3}\cline{4-5}\cline{6-7}
 & PBE & LDA & PBE & LDA & PBE & LDA \\ \hline
a ($\text{\AA}$)   & 5.082 & 4.994 & 3.594 & 3.533 & 5.145 & 5.048  \\
b ($\text{\AA}$)   & --   & --   & --   & --   & 5.206 & 5.142  \\ 
c ($\text{\AA}$)   & --   & --   & 5.225 & 5.076 & 5.326 & 5.206  \\
$\beta$ ($^\circ$) & --   & --   & --   & --   & 99.63 & 99.53  \\
\end{tabular}
\end{ruledtabular}
\caption{\label{tab:properties}Structural parameters of cubic (spacegroup $Fm3m$), tetragonal (spacegroup $P4_2/nmc$) and monoclinic (spacegroup $P2_1/c$) phases of HfO$_2$.}
\end{table*}

\newpage

\appendix

\section{Supplementary material}

\paragraph*{Structural parameters}

All our calculations are based on density functional theory within local density approximation (LDA) or generalized gradient approximation (GGA) with the parametrization of Perdew-Burke-Ernzerhof (PBE) in the framework of plane wave projector-augmented wave (PAW) method as implemented in the \textsc{vasp} code.\cite{Kresse_PRB_1996, Kresse_PRB_1999} The DFT calculations are performed using the version 5.4 of LDA and PBE PAW potentials of \textsc{vasp} (Hf\_sv\_GW and O\_GW for Hf and O, respectively).

We study cubic (spacegroup $Fm3m$), tetragonal (spacegroup $P4_2/nmc$) and monoclinic (spacegroup $P2_1/c$) phases of HfO$_2$. Each structure is relaxed until the maximum residual forces are less than $10^{-3}$~eV/$\text{\AA}$. For the relaxation we use a fine $\mathbf{k}$ mesh of $12 \times 12 \times 12$, $12 \times 12 \times 8$ and $8 \times 8 \times 8$ for cubic, tetragonal and monoclinic phases, respectively. Calculated structural parameters are summarized in Table~\ref{tab:properties}. 

\begin{figure}[b!]
\includegraphics[width=\columnwidth]{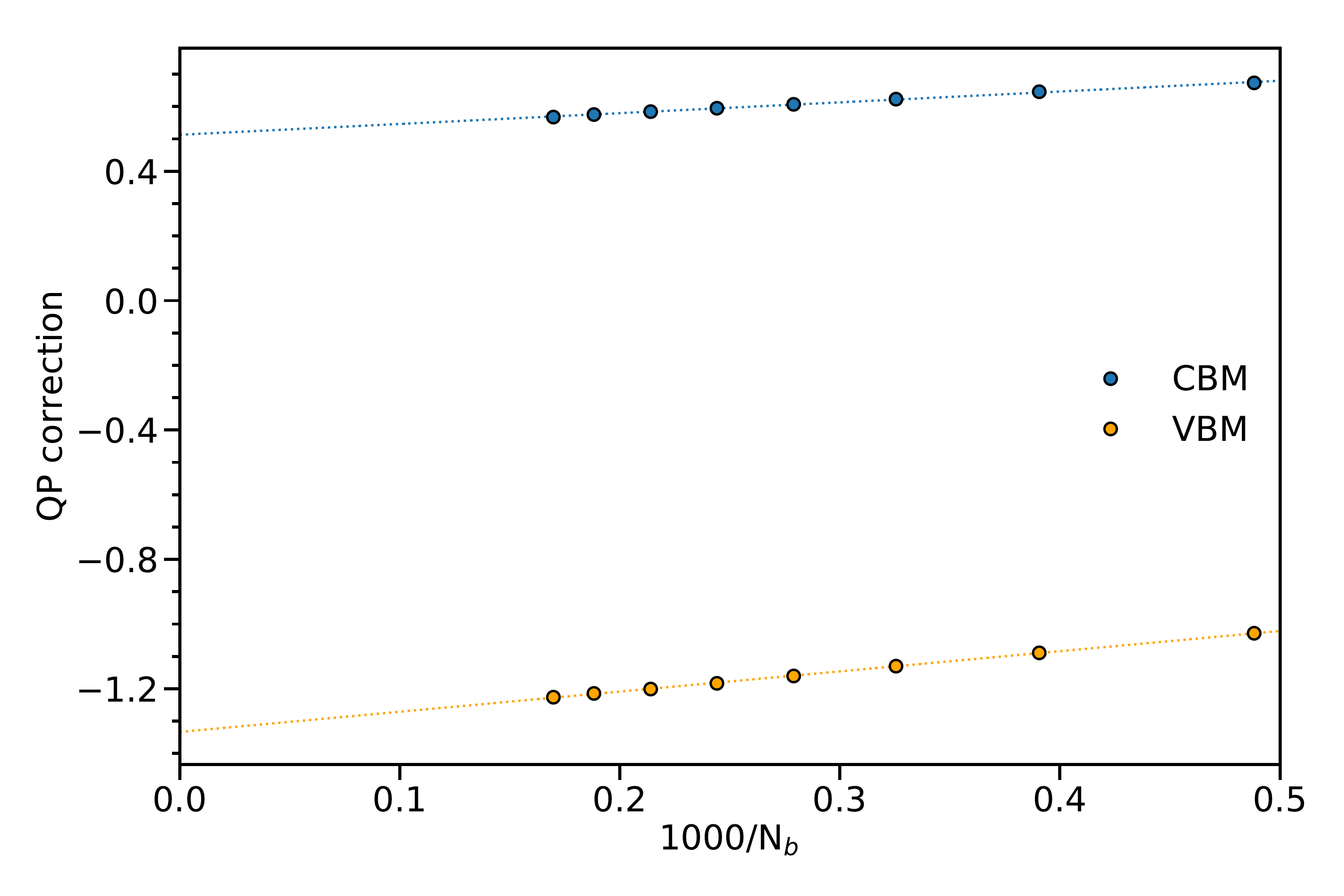}
\caption{QP corrections of the Kohn-Sham eigenvalues as a function of the inverse of the number of empty bands N$_b$ (or number of plane waves N$_\text{pw}$). The dotted lines show the linear extrapolation of QP corrections (see text for details).}
\label{fig:gw_Nb_convergence} 
\end{figure}

\paragraph*{One-shot $G_0W_0$ convergence}
\label{sec:g0w0}

We carefully examine the convergence of our $GW$ calculations in order to achieve QP band gap values converged within 0.1~eV.
Indeed, to calculate the  response function and the correlation part of self-energy, a summation over empty states is required and quasi particle (QP) energies exhibit a very slow convergence with respect to the number of virtual orbitals.
\begin{figure}[b!]
\includegraphics[width=\columnwidth]{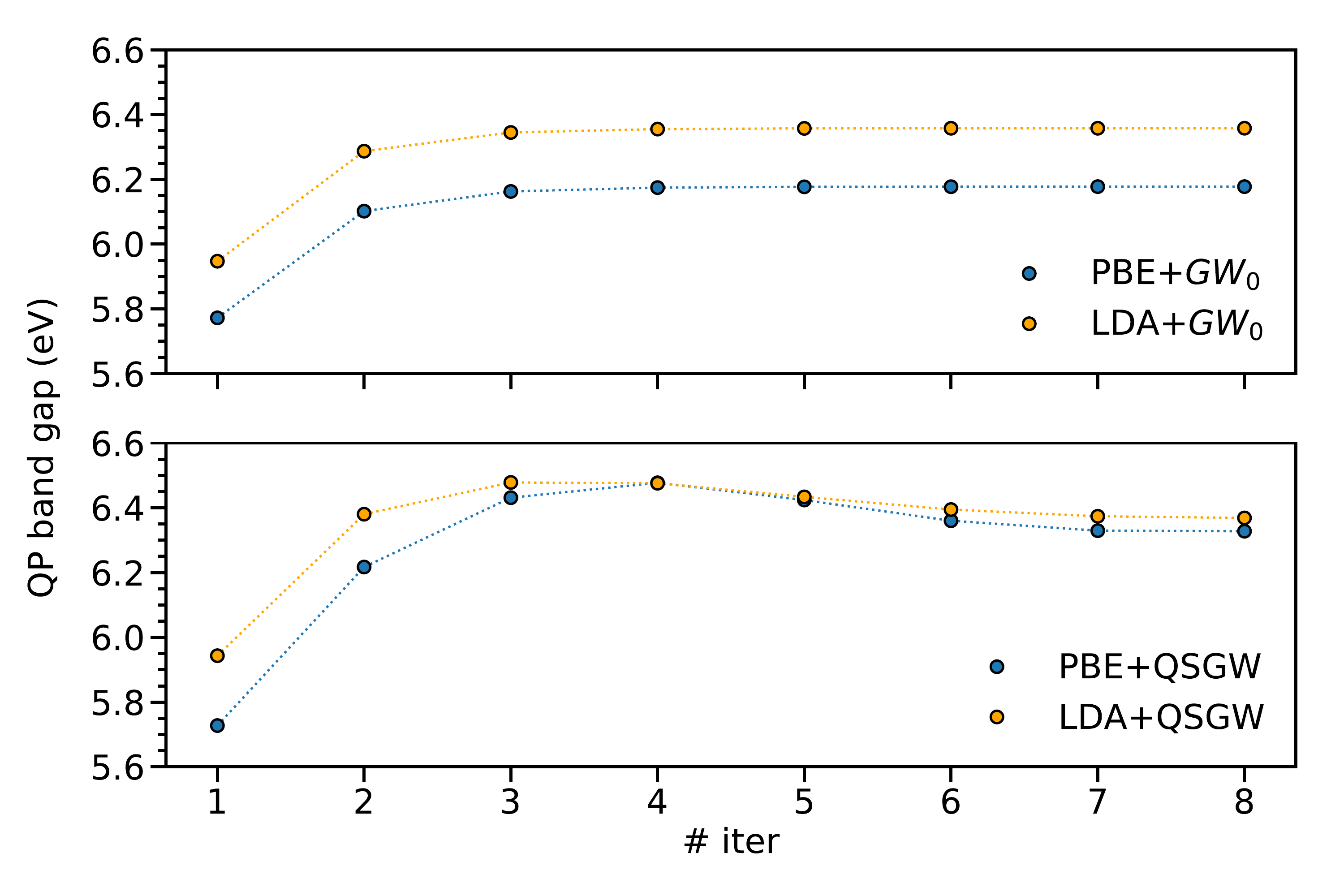}
\caption{Convergence of QP band gap as a function of self-consistent $GW$ iterations at $GW_0$ (top panel) and QS$GW$ (bottom panel) level.}
\label{fig:gw_convergence} 
\end{figure}
\begin{table*}[t!]
\renewcommand{\tabcolsep}{2mm}
\begin{ruledtabular}
\begin{tabular}{llccccccc}
  & & \multicolumn{2}{c}{DFT}     & \multicolumn{3}{c}{Convergence parameters} & \multicolumn{2}{c}{QP energies} \\ \cline{3-4}\cline{5-7}\cline{8-9}
  & & xc functional & core elect. & $\mathbf{k}$ mesh & N$_b$ & Dyn. Scr. & $E_{g}$ (eV) & $\Delta E_{g}^{GW}$ (eV) \\ \hline
\multirow{4}{*}{This work} %
  & \multirow{2}{*}{$G_0W_0$} & PBE & \multirow{2}{*}{PAW} & \multirow{2}{*}{$4 \times 4 \times 4$} & \multirow{2}{*}{2000} & \multirow{2}{*}{Full Freq.} & 5.77 &  1.76 \\
                     &  & LDA & & & & & 5.95 & 2.02 \\
  & \multirow{2}{*}{$GW_0$} & PBE & \multirow{2}{*}{PAW} & \multirow{2}{*}{$4 \times 4 \times 4$} & \multirow{2}{*}{2000} & \multirow{2}{*}{Full Freq.} & 6.18 & 2.17 \\
                     &  & LDA & & & & & 6.36 & 2.43 \\
\multirow{2}{*}{Ref.~\onlinecite{Gruning_PRB_2010,Gruning_private_2018}} %
  & $G_0W_0$ & \multirow{2}{*}{LDA} & \multirow{2}{*}{NCPP} & \multirow{2}{*}{$4 \times 4 \times 4$} & \multirow{2}{*}{600} & \multirow{2}{*}{GN-PPM} & 5.7 & 1.9\\
                     & $GW_0$ & & & & & & 5.9 & 2.1\\
\multirow{2}{*}{Ref.~\onlinecite{Jiang_PRB_2010}} %
  & $G_0W_0$ & \multirow{2}{*}{LDA} & \multirow{2}{*}{FP-LAPW} & \multirow{2}{*}{$2 \times 2 \times 2$} & \multirow{2}{*}{---} & \multirow{2}{*}{Analytical cont.} & 5.45 & 1.50 \\
                     & $GW_0$ & & & & & & 5.78 & 1.83 \\
\end{tabular}
\end{ruledtabular}
\caption{\label{tab:gw_compare}Comparison of QP band gaps ($E_{g}$) and $GW$ corrections ($\Delta E_g^{GW} = E_g^{GW} - E_g^\mathrm{DFT}$) from different $GW$ calculations on monoclinic HfO$_2$. The calculations differ by the level of self--consistency, starting mean-field theory and $GW$ convergence parameters. The convergence parameters include: the $\mathbf{k}$ mesh, the number of empty bands (N$_b$) and the method to describe dynamical screening (Dyn. Scr.).}
\end{table*}
Furthermore, the number of empty bands N$_b$, the corresponding orbital basis set N$_\text{pw}$ (controlled by the plane waves cutoff E$_\text{pw}$) and the size of the response function basis set  N$_\text{pw}^\chi$ (controlled by the plane waves cutoff E$_\text{pw}^\chi$) have to be increased simultaneously~\cite{Klimes_2014}. Therefore, in our convergence study we fix $E_\text{pw}^\chi = 2/3 E_\text{pw}$ and use the complete plane waves basis set (i.e.\ N$_b$ = N$_\text{pw}$) for each considered E$_\text{pw}$. We consider the monoclinic phase of HfO$_2$ with $E_\text{pw}$ ranging from 350~eV to 700~eV. A frequency grid with 100 frequency points is used to represent the polarizability and a $4 \times 4 \times 4$ $\Gamma$-centered \textbf{k}-mesh is used to sample the Brillouin  zone. Fig.~\ref{fig:gw_Nb_convergence} shows the QP corrections ($\Delta \epsilon$) of the Kohn-Sham eigenvalues as a function of 1/N$_b$ (which equals 1/N$_\text{pw}$) for a $G_0W_0$ calculation on top of PBE (a similar behavior is observed for an LDA starting point). The dotted lines show the linear extrapolation of QP corrections ($\Delta \epsilon(N_b) = A/N_b + \Delta \epsilon (\infty)$) where only the values corresponding to $E_\text{pw} \geq 500$~eV are included in the fit. The extrapolated QP corrections of the valence band maximum (VBM) at $\Gamma$ and conduction band minimum (CBM) at $B$ are respectively -1.33~eV and 0.51~eV giving an extrapolated band gap of 5.85~eV.
Our convergence study suggests that a 500~eV plane waves cutoff (3603 plane waves) allows to achieve a band gap converged within 80~meV.
For this basis set, the number of empty bands can be decreased to 2000 without deteriorating the convergence (see bottom right panel of Fig.~\ref{band-plot} in the main text). 

We also check the influence of the number of frequency points and \textbf{k} mesh. We find that increasing the frequency grid from 100 to 200 points  only changes the QP band gap by 15~meV but tend to compensate the error done due to incompleteness of the plane waves basis set. When using a  finer k-point sampling of $6 \times 6 \times 6$, change in the QP band gap is below 1~meV.

Our convergence study shows that the numerical convergence of our calculated QP band gaps is below 80~meV. In table~\ref{tab:gw_compare}, we summarize the convergence parameters used in our work with those from previous theoretical studies.

\begin{figure}[b]
\includegraphics[width=\columnwidth]{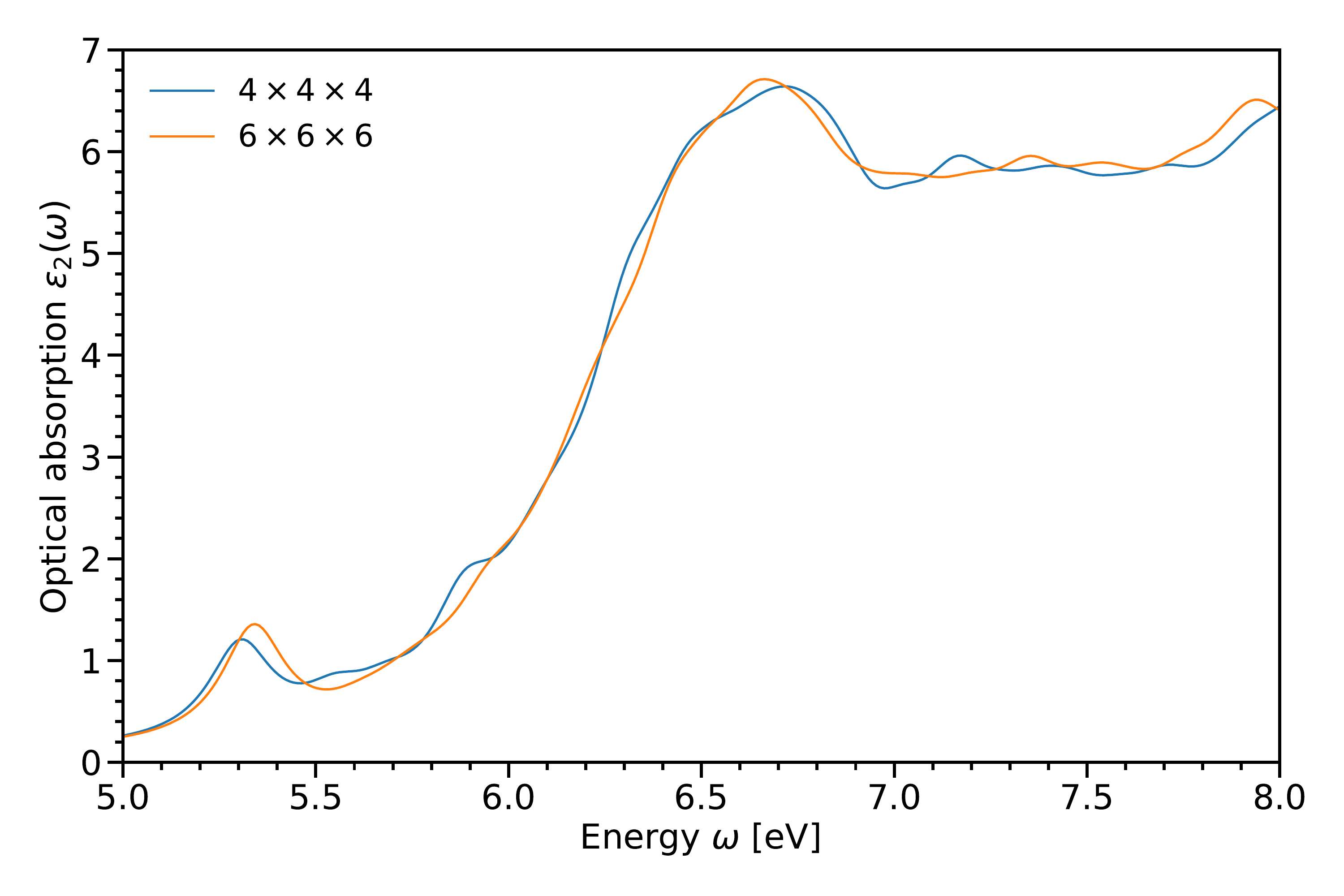}
\caption{Imaginary part of the dielectric function $\varepsilon_2$ for the monoclinic phase of HfO$_2$ calculated on top of PBE+G$_0$W$_0$ for $4 \times 4 \times 4$ and $6 \times 6 \times 6$ $\mathbf{k}$ mesh.  A Gaussian broadening of 0.1~eV is used.}
\label{fig:bse_convergence} 
\end{figure}

\paragraph*{Self-consistent $GW$ convergence}

Self-consistent $GW$ calculations are carried out with the same parameters as for $G_0W_0$. For $GW_0$ and QS$GW$ calculations, we include respectively 256 and 512 QP energies in the self-consistent procedure for monoclinic HfO$_2$. In the case of $GW_0$, 4 self-consistent iterations are enough to get converged band gap. For QS$GW$, 8 self-consistent $GW$ iterations allowed to get converged QP energies and band gaps within 5~meV as shown in Fig.~\ref{fig:gw_convergence}.

\begin{figure}[b]
\includegraphics[width=\columnwidth]{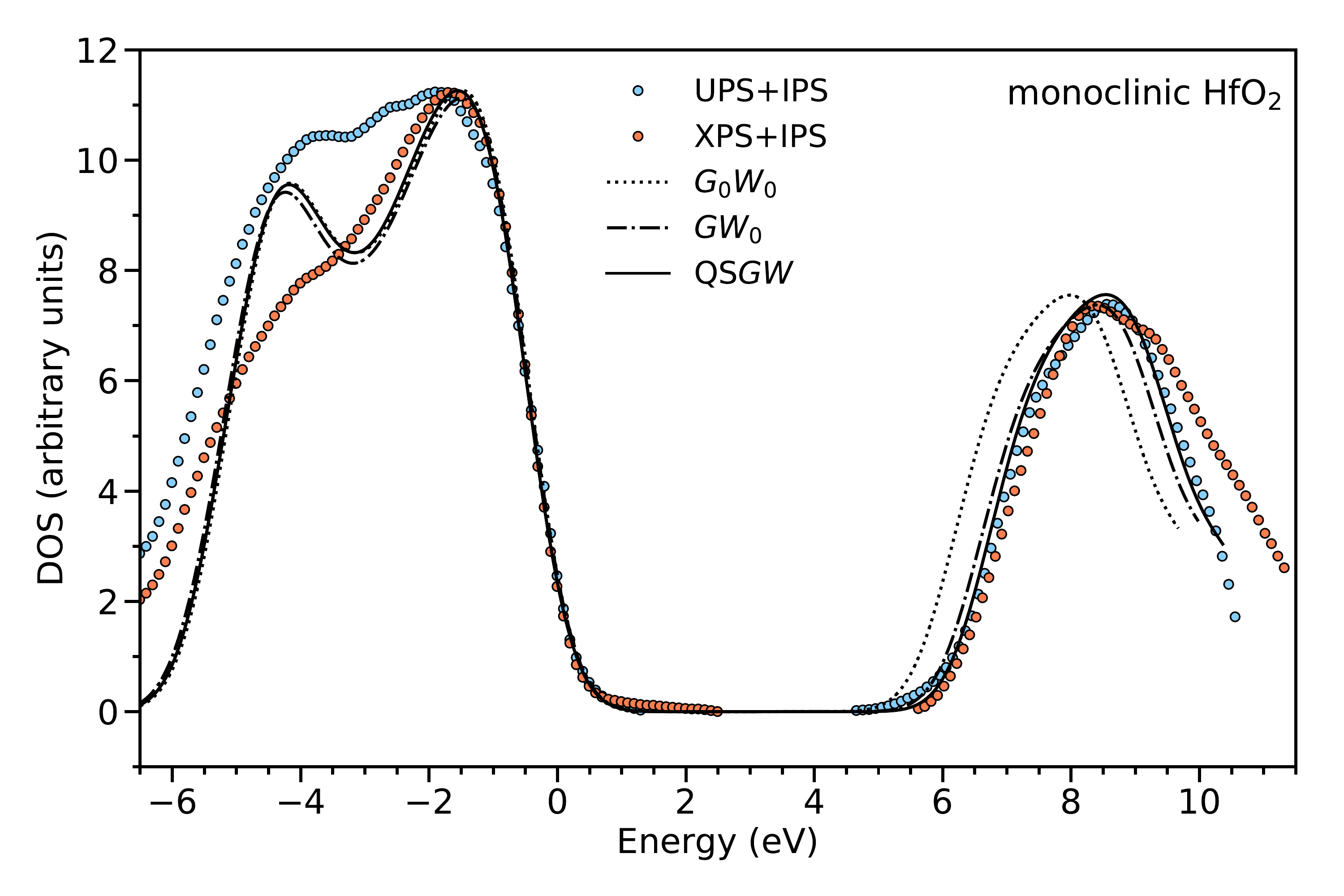}
\caption{Density of states (DOS) of monoclinic HfO$_2$ in the $G_0W_0$, $GW_0$, and QS$GW$ approximations compared to UPS+IPS\cite{Bersch_PRB_2008} and XPS+IPS\cite{Sayan_JAP_2004} spectra.
The theoretical DOS have been interpolated using MLWF on a $40 \times 40 \times 40$ k-point grid and convoluted with a gaussian broadening of 0.7~eV.
Experimental spectra have been aligned at the valence band-edge (the zero of the energy for all theoretical DOS), and the conduction IPS and valence PS independent measurements have been rescaled separately to match the height of the theoretical DOS.
}
\label{fig:dos} 
\end{figure}

\paragraph*{Bethe-Salpeter equation (BSE) convergence}

The excitonic properties are determined by solving the Bethe-Salpeter equation (BSE) within the Tamm-Dancoff approximation on top of the $GW$ quasiparticle band structure. BSE calculations usually require fine \textbf{k}-point sampling to converge exciton spectra. However such calculations are computationally very expensive for both $GW$ and BSE. In the case of monoclinic HfO$_2$ we test the \textbf{k}-point convergence of the BSE calculation on top of $G_0W_0$ using $4 \times 4 \times 4$ and $6 \times 6 \times 6$ grids. Fig.~\ref{fig:bse_convergence} shows the imaginary part of the dielectric function $\varepsilon_2$ (we ignore the polarization dependence and assume $\varepsilon = \left( \varepsilon_{xx} + \varepsilon_{yy} + \varepsilon_{zz} \right) / 3$). A simple visual inspection suggests that the two \textbf{k}-meshes give very similar spectra. More quantitatively, the exciton binding energies of $4 \times 4 \times 4$ and $6 \times 6 \times 6$ meshes are 0.56~eV and 0.52~eV, respectively.

A more complete convergence study with respect to \textbf{k}-mesh could be carried out using a model BSE (mBSE) scheme or interpolation techniques but have not been considered in this work.

\paragraph*{DOS}

For monoclinic HfO$_2$, we present in Fig.~\ref{fig:dos} the density of states (DOS) calculated in the $G_0W_0$, $GW_0$, and QS$GW$ approximations compared to the experimental UPS+IPS\cite{Bersch_PRB_2008} and XPS+IPS\cite{Sayan_JAP_2004} spectra.
The figure shows that the theoretical underestimation of the band gap decreases from $G_0W_0$, passing by $GW_0$, to QS$GW$.

\bibliography{biblio}

\end{document}